\documentclass[pre,twocolumn,amsmath,showpacs,floatfix]{revtex4}
\usepackage{color,graphicx}
\begin{document}

\title{Pulse propagation in a chain of o-rings with and without precompression}
\author{Italo'Ivo Lima Dias Pinto}
\author{Alexandre Rosas}
\email{arosas@fisica.ufpb.br}
\affiliation{Departamento de F\'{\i}sica, CCEN, Universidade Federal da Para\'{\i}ba, Caixa Postal 5008, 58059-900, Jo\~ao Pessoa, Brazil.}
\author{Aldo H. Romero}
\affiliation{Cinvestav-Quer\'etaro, Libramiento Norponiente 200, 76230, Fracc. Real de Juriquilla, Quer\'etaro, Quer\'etaro, M\'exico.}
\author{Katja Lindenberg}
\affiliation{Department of Chemistry and Biochemistry and BioBircuits Institute, University of California San Diego, La Jolla, California 92093-0340, USA.}

\begin{abstract}
We implement a binary collision approximation to study pulse propagation in a chain of o-rings.
In particular, we arrive at analytic results from which the pulse velocity is obtained by simple
quadrature.  The predicted pulse velocity is compared to the velocity
obtained from the far more resource-intensive numerical integration of
the equations of motion. We study chains without precompression, chains precompressed
by a constant force at the chain ends (constant precompression), and chains precompressed
by gravity (variable precompression).
The application of the binary collision approximation to
precompressed chains provides an important generalization of a successful theory that had up to this
point only been implemented to chains without precompression, that is, to chains in a sonic vacuum.
\end{abstract}


\maketitle 

\section{Introduction}
The dynamics of pulse propagation in granular chains continues to be of keen interest
to physicists and engineers both because of the theoretical challenges that it poses,
and because of the prospect of related technological applications. Among other applications,
it has been argued
that understanding of pulse propagation in granular media may lead to the development
of new shock absorbers and of instruments for detection of
buried objects~\cite{sen_solitonlike_1998}. Most recently, a powerful new nonlinear acoustic lens
consisting of ordered arrays of granular chains has been shown to be capable of generating
high-energy acoustic pulses that can be used in a large variety of applications from biomedical
imaging to defense systems and damage detection in materials~\cite{daraio_soundbullets_2010}.
While these applications involve propagation in higher
dimensional granular realms, a thorough understanding of the dynamics in one dimension is often
pursued as a way to lay the groundwork for the problem, or even because the systems are actually
composed of arrays of granular chains.  Theoretically,
a granular chain appears simple, and yet its discrete non-linear character
immediately imposes severe difficulties on the development of analytic methods to
explain and predict the behavior of energy pulses in even these simplest granular environments.
As a result, the available information is obtained mostly from
numerical~\cite{sinkovits_nonlinear_1995,xu_power-law_2001,herbold_solitary_2007,daraio_energy_2006,rosas_observation_2007,rosas_short-pulse_2008}
and occasionally from experimental~\cite{nakagawa_impulse_2003,coste_solitary_1997} results. The numerical
approaches are notoriously resource-intensive, and the experiments are constrained by the particular
experimental setup.

The numerical and experimental results have been supplemented with some analytic results that have
proven extremely helpful and accurate.  The accuracy is particularly interesting in 
view of the fact that the two most successful theories
approach the problem from two essentially orthogonal points of view, both
of which have provided excellent results.  Historically, the first was based on a long-wavelength
approximation~\cite{nesterenko_propagation_1983,lazaridi_observation_1985} that assumes that a chain
can be treated as a continuum augmented with first-order corrections due to the discreteness
of the system.  This approach was applied to granular chains in which the granules just touch, so
that there are no intergranular gaps but also no initial precompression. The long-wavelength
solution to this problem brought to light the existence of solitary waves in granular chains.
This purely nonlinear scenario was dubbed a ``sonic vacuum" because it supports no
sound waves.  The continuum approach has been applied in the
absence~\cite{nesterenko_propagation_1983,lazaridi_observation_1985,nesterenko_dynamics_2001} as
well as in the presence of dissipation~\cite{rosas_pulse_2003}. 
The success of the continuum approach is particularly noteworthy because it presumes
a pulse width that is large compared to the size of the granules, and yet the observed pulse
as well as the solitary wave solution that emerges from this approach extend over only a few
granules.  The continuum methodology has been applied to a variety of chains with different granular
configurations and initial conditions~\cite{nesterenko_dynamics_2001},
but the mathematics is quite cumbersome. 

A second alternative approach developed more recently adopts an orthogonal point of view in that
it is based on a binary collision approximation that presumes that intergranular collisions involve
only two granules at a time. This extremely-short-wavelength approach has in fact been shown to
yield results that are even closer to numerical simulation results for the ``canonical" case of
a chain of spherical granules with initially no gaps and no
precompression~\cite{rosas_pulse_2004-1}. Not only does this method yield highly accurate results
for that case, but it has been successfully applied to a number of other configurations that are not
easily accessible to the continuum approach, namely, to
tapered~\cite{harbola_pulse_2009}, decorated~\cite{harbola_pulse_2009-1}, and randomly
decorated chains~\cite{harbola_pulse_2010} and to one-dimensional granular
gases~\cite{pinto_energy_2009}. However, it has not yet been applied to any kind of
precompressed chain. 

In this contribution, we study pulse propagation in a granular chain of toroidal rings (o-rings)
placed between rigid cylinders that act as nonlinear springs. For such granules,
the purely repulsive force characteristic of dry granular materials is modified  by
the topological properties of the o-rings, leading to a hard potential proportional
to the seventh power of the compression in addition to the usual softer Hertz potential
proportional to the compression to power $5/2$. We are directed to this particular system
because it has been studied both numerically and experimentally~\cite{herbold_solitary_2007}.
Here, we make use of the binary collision approximation to study the pulse velocity in
such chains of o-rings, extending the approximation not only to a new geometry but also
beyond the case of a sonic vacuum to precompressed chains.  The extension of the binary collision
approximation to precompressed chains provides an important generalization of a successful theory
that had up to this point only been implemented for chains without precompression.

The paper is organized  as follows. In Sec.~\ref{sec:model} we present the equations of
motion for the granular chain of o-rings and we describe the binary collision approximation.
Next, we compare the results of this approach with those of the numerical integration of
the equations of motion in Sec.~\ref{sec:numerical}. Finally, in Sec.~\ref{sec:conclusion} we
briefly summarize our results. 

\section{The model} 
\label{sec:model}
Consider a chain of $N$ equal o-rings, hereafter also called the ``granules." The
double power law character of the elastic
interaction of the o-rings leads to the following equation of motion for 
granule $k$ in the interior of the chain,
\begin{equation}
\begin{split}
\frac{d{y}_k^2}{d\tau^2} =& A [ (y_{k-1} - y_k)^{3/2} - (y_{k} - y_{k+1})^{3/2} ] \\
&+ B [ (y_{k-1} - y_k)^6 - (y_{k} - y_{k+1})^6 ] + F_k/m,
\label{eq:motion}
\end{split}
\end{equation}
where $y_k$ is the displacement of the $k$-th granule at time $\tau$ from its equilibrium
position. Granule $k$ is subject to the external force $F_k$, and $m$ is the mass of the
cylinders separating the o-rings (which is much larger than the mass of the o-rings). 
The constants $A = 1.25 \pi D E/ m d^{1/2}$ and $B = 50 \pi D E/ m d^5$ are constants that
characterize the elastic properties of
the material~\cite{herbold_solitary_2007}, where $d$
and $D$ are respectively the cross-section and mean diameter of the o-ring, and $E$ is
the Young's modulus of the o-rings.  For comparison with experiments, we present
the values for these constants for the teflon o-rings used
in~\cite{herbold_solitary_2007}: $E =1.46 GPa, \; D = 7.12 mm, \; d = 1.76 mm$ and
$m = 3.276 g$.

The granules at the ends of the chain must be considered separately. While the leftmost
granule does not have any granule pushing it to the right, a constant force $F_1$ may
be applied in this direction in the case of precompression. Therefore, its equation of motion reads
\begin{equation}
\frac{d{y}_1^2}{d\tau^2} = F_1/m - A (y_{1} - y_{2})^{3/2} - B (y_{1} - y_{2})^6.
\label{eq:leftmost}
\end{equation}
Similarly, the rightmost granule moves according to
\begin{equation}
\frac{d{y}_N^2}{d\tau^2} = - F_N/m + A (y_{N-1} - y_{N})^{3/2} + B (y_{N-1} - y_{N})^6.
\label{eq:rightmost}
\end{equation}
Without precompression, $F_k = 0$ for all granules, while for a constant precompression,
$F_1 = -F_N = F$, and $F_k = 0$ for all other granules. In the case of
a vertical chain subject to gravity with labels running downward,
$F_k = k m g$ for $k=1, 2, \dots, N-1$ and $F_N = - N m g$,
$g$ being the acceleration due to gravity.
Since we are interested in pulse propagation, initially all the granules are at rest
except for the leftmost or top granule, which has an initial velocity $v_0$.
In the case of precompression, either by an external constant driving force or
by gravity, the initial positions of the granules are the equilibrium positions. 
For a chain in a sonic vacuum (that is, without precompression), initially the
granules just touch each other.

We proceed by defining the scaled variables~\cite{rosas_dynamics_2003}
\begin{equation}
  y_k = \left ( \frac{v_0^2}{A} \right )^{2/5} x_k, \quad \tau = \frac{1}{v_0} \left ( \frac{v_0^2}{A} \right )^{2/5} t_k,
\label{eq:scaled}
\end{equation}
in terms of which the equations of motion are written as
\begin{eqnarray}
\ddot{x}_1 &=& f_1 - (x_{1} - x_{2})^{3/2} - b (x_{1} - x_{2})^6, \nonumber\\
\ddot{x}_k &=& f_k + [ (x_{k-1} - x_k)^{3/2} - (x_{k} - x_{k+1})^{3/2} ] \nonumber\\
&+& b [ (x_{k-1} - x_k)^6 - (x_{k} - x_{k+1})^6 ], \label{eq:motion-scaled} \\
\ddot{x}_N &=& - f_N + (x_{N-1} - x_{N})^{3/2} + b (x_{N-1} - x_{N})^6,\nonumber
\end{eqnarray}
where a dot denotes a derivative with respect to $t$ and where we have defined
\begin{equation}
  b = \frac{B}{A}\left ( \frac{v_0^2}{A} \right )^{9/5}, \quad f_k = \frac{F_k}{mv_0^2} \left ( \frac{v_0^2}{A} \right )^{2/5}.
\end{equation}
Consequently, the parameter $b$ measures the relative strength of the two power law terms
vs the initial velocity, and $f_k$ plays the role of the external force.

\subsection{Binary collision approximation} 
\label{sub:binary}
The binary collision approximation is based on the assumption that the pulse propagates
through the chain by a sequence of binary collisions.
Since this is not exactly the case, we must also specify the moment of passage of the
pulse from one granule to the next.
We say that the pulse moves from granule $k$ to granule $k+1$ when the velocity of the
latter surpasses the velocity of the former. With this approximation, instead of
having to take into account all the equations of motion at once, we can focus on the
interaction of just two granules at a time. Since the granules in the original chain are initially
in equilibrium, so are the two granules in the binary collision approximation.
Consequently, the equations of motion of these two particles during the collision may be written as
\begin{eqnarray}
  \ddot{x}_k &=& f - (x_{k} - x_{k+1})^{3/2} - b (x_{1} - x_{2})^6, \\ 
  \ddot{x}_{k+1} &=& - f + (x_{k} - x_{k+1})^{3/2} + b (x_{1} - x_{2})^6,
\end{eqnarray}
where $f$ is the force causing the compression of the two granules at the beginning of the
collision. In the case of a sonic vacuum, this force is zero.
When a constant external compression force is applied at the ends of the chain, 
the force is positive and $f$ equals this external force for each pair of granules.
In the presence of a gravitational force 
it equals $k \cal{G}$, the scaled gravitational constant $\cal{G}$ being related to the
unscaled gravitational constant $g$ by
\begin{equation}
{\cal{G}} = \frac{g}{v_0^2}\left(\frac{v_0^2}{A}\right)^{2/5}.
\label{gravityscaling}
\end{equation}

We define $z = x_k - x_{k+1}$ and, for simplicity of notation,
for now we drop the $k$ subscripts on $z$ and $f$.
Subtracting the two equations, we have
\begin{equation}
  \ddot{z} = 2 f - 2 z^{3/2} - 2 b z^6,
\end{equation}
which describes the motion of a fictitious particle of unit mass whose 
displacement is $z$ and which is moving in the potential
\begin{equation}
  V(z) = - 2 f z + \frac{4}{5} z^{5/2} + \frac{2}{7} b z^7.
\label{eq:potential}
\end{equation}
Next we make use of conservation of energy to write
\begin{equation}
  \frac{1}{2} \dot{z}^2 + V(z) = \frac{1}{2} \dot{z}_0^2 + V(z_0),
\label{eq:conservation}
\end{equation}
where $\dot{z}_0 = 1$ is the velocity of the incoming granule at the beginning of the
collision, and $z_0$ is obtained from the equilibrium condition
\begin{equation}
    f = z_0^{3/2} + b z_0^6.
\label{eq:z0}
\end{equation}
As discussed above, the pulse is said to reside on
granule $k$ until the velocities of the two granules become equal. At that moment
$\dot{z}=0$, the pulse is said to move onto the next granule, and the compression
is maximum. Hence, the maximum compression $z_m$ is obtained from the energy
conservation condition Eq.~(\ref{eq:conservation}) as the solution of
\begin{equation}
  \frac{8}{5} \left ( z_m^{5/2} - z_0^{5/2}\right )+ \frac{4}{7} b
\left ( z_m^7 - z_0^7 \right ) - 4 f \left ( z_m - z_0 \right ) =  1.
\label{eq:zm}
\end{equation}
Once we know the initial and maximum compression, the residence time, i.e., the time
spent by the pulse on a given granule, may be obtained as
\begin{equation}
  T_k = \int_{z_{0,k}}^{z_{m,k}} \frac{dz}{\dot{z}} = \int_{z_{0,k}}^{z_{m,k}}
\frac{dz}{\sqrt{1 + V(z_{0,k}) - V(z)}},
\label{eq:residence}
\end{equation}
where we have made use of Eq.~(\ref{eq:conservation}) and reinstated the subscript $k$.

In summary, in order to calculate the pulse velocity as it passes through
granule $k$, we need to find the solutions $z_0$ and $z_m$ of Eqs. (\ref{eq:z0})
and (\ref{eq:zm}), respectively, and then we must numerically integrate
Eq. (\ref{eq:residence}) (since analytic integration appears impossible) to find the
residence time. The pulse velocity is the inverse of the residence time, 
\begin{equation}
  c_k = 1/T_k.
\label{eq:pulsevel}
\end{equation}

\section{Numerical results} 
\label{sec:numerical}
In this section we compare the pulse velocity predicted by the binary
collision approximation, Eq.~(\ref{eq:pulsevel}), with the results of the numerical
integration of the equations of motion, Eqs.~(\ref{eq:motion-scaled}). We consider three
cases: chains without precompression, chains with precompression caused by a
constant force at the edges of the chain, and chains with precompression caused by gravity.

\subsection{Chains without precompression} 
\label{sub:nopre}

This is the simplest case. In the absence of precompression, $z_0$ and $f$ vanish.
The $k$-independent potential then becomes
\begin{equation}
  V(z) = \frac{4}{5} z^{5/2} + \frac{2}{7} b z^7,
\label{eq:potential-nopre}
\end{equation}
and the equation for the maximum compression is simplified to
\begin{equation}
  \frac{8}{5} z_m^{5/2} + \frac{4}{7} b z_m^7 = 1.
\label{eq:zm-nopre}
\end{equation}
These simplifications lead to the residence time 
\begin{equation}
  T = \int_{0}^{z_m} \frac{dz}{\sqrt{1 - \frac{8}{5} z_m^{5/2} - \frac{4}{7} b z_m^7}}.
\label{eq:residence-nopre}
\end{equation}
In the limit of small and large $b$, one or the other of the two terms
in the potential may be neglected. In these limits $z_m$ can be calculated exactly
from Eq.~(\ref{eq:zm-nopre}). Furthermore, the integral (\ref{eq:residence-nopre})
can then be performed exactly~\cite{rosas_pulse_2004-1}, leading to the low-$b$ pulse velocity 
 \begin{equation}
   c = \frac{2^{1/5} 5^{3/5} \Gamma (9/10)}{\sqrt{\pi} \Gamma (2/5)} \simeq 0.820,
\label{eq:small-b}
 \end{equation}
and the large-$b$ velocity
 \begin{equation}
   c = \frac{2^{2/7} 7^{6/7} b^{1/7} \Gamma (9/14)}{\sqrt{\pi} \Gamma (1/7)} \simeq 0.779 b^{1/7}.
\label{eq:large-b}
 \end{equation}

\begin{figure}
 \includegraphics[angle=-90,width=8cm]{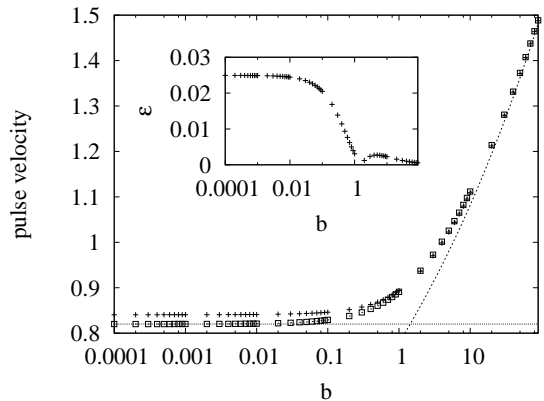}
 \caption{Pulse velocity as a function of the parameter $b$ in the absence of
precompression. In the main figure, the plus signs refer to the numerical simulation
results and the open squares to the binary collision approximation. The dotted line is
the plot of Eq.~(\ref{eq:small-b}) and the dashed line is the plot of Eq.~(\ref{eq:large-b}).
The inset shows the relative error $\varepsilon$. \label{fig:nopre}}
\end{figure}

In Fig.~\ref{fig:nopre} we show the pulse velocity as a function of the parameter $b$.
This parameter measures the relative weight of the two terms in the granular interaction.
For small values of $b$, the interaction is almost Hertzian while for large values
the $z^6$ force is dominant. This behavior is clearly illustrated by the dotted
and dashed lines, which are the plots of Eqs.~(\ref{eq:small-b}) and (\ref{eq:large-b}),
respectively. The excellent agreement of the prediction of the binary
collision approximation is evident over the entire range of values of $b$. In the inset
we show the relative error $ \varepsilon = \left|(c_b - c_n)/c_b\right|$ between
the pulse velocity predicted by the binary collision approximation, which we call $c_b$,
and the value obtained from the numerical integration of the equations of motion, which
we call $c_n$. The plot shows that $ \varepsilon $ is always smaller than 3\%.
Furthermore, as $b$ increases the error decreases because the harder potential becomes more
important and the associated pulse is narrower. Consequently the binary collision approximation
turns out to be increasingly more precise, that is, the idea that only two granules participate
in each collision becomes increasingly more correct~\cite{rosas_pulse_2004-1}.

We end our discussion of uncompressed chains with a comparison of
our results with the experiments of Herbold and Nesterenko~\cite{herbold_solitary_2007}. In
their work, the pulse was generated by the impact of a stainless steel sphere (not part of the
chain) of $0.455 g$ of mass with velocities ranging from $1 m/s$ to $800 m/s$. In our model
the pulse is generated by the first granule of the chain, whose mass is $3.276 g$ in their
experiment. We adjust our initial momentum transfered to the chain to theirs by choosing
the parameter $b$ appropriately, in the range $[10^{-10}, 5]$. At the lower
limit the dominant potential term is the Hertzian, and the pulse velocity is
$c = 0.82$ (see Fig~\ref{fig:nopre}), which translates to a pulse velocity in physical units of
around $250 m/s$. The difference between our results and the experimental ones in this regime
is about $172 m/s$ and is primarily due to the difference in the impulse generating method.
However, it is reassuring that our results are within a factor of 2
of the experimental results. Experimental results were not shown for the large $b$ limit
so we are not able to compare with our theory.  Our binary collision approximation
predicts that the pulse velocity in this regime should be around $1100 m/s$.

\subsection{Constant precompression} 
\label{sub:constpre}
\begin{figure}
 \includegraphics[width=8cm]{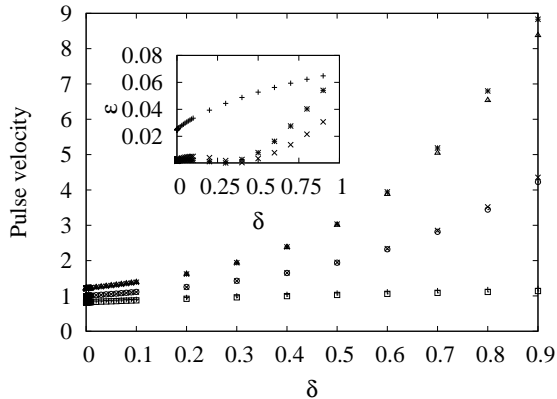}
 \caption{Pulse velocity as a function of the compression. In the main figure, the
plus signs (open squares), crosses (open circles) and stars (triangles) refer to the
binary collision approximation
for $b = 10^{-5}, \;4$ and $20$, respectively. The inset shows the relative error
$\varepsilon$ for the same values of $b$. \label{fig:constpre}}
\end{figure}

When we apply a constant force at the ends of the chain, thus pressing the granules together,
the precompression of any pair of granules is the same, $z_0 = \delta$, as is the
force $f$ on each granule. Therefore, the pulse again travels with a constant ($k$-independent) velocity
along the chain. Figure~\ref{fig:constpre} shows that this constant pulse speed increases
as the precompression (or, alternatively, the force) increases. The inset shows the relative
error, which increases with increasing precompression. This error increase is a reflection of the
fact that as precompression increases, more than two granules become actively engaged in any
collision event.
The force caused by granule $k+2$ on granule $k+1$ for a power law
potential proportional to $z^n$ is $(n-1) \left (\delta_0/ \delta_z \right )$,
where $ \delta_0$ is the precompression between granules $k+1$ and $k+2$ and $ \delta_z$ is
the additional compression between them caused by the traveling pulse. Hence, for
large $ \delta_0$, this interaction
is enhanced and the binary collision approximation is less effective. The inset
also shows that for
larger values of $b$, the relative error $\varepsilon$ is initially smaller, as discussed in the
previous section, but as $ \delta =\delta_0+\delta_z$ increases, the increase in the
relative error is more prominent. Further, the increase in $\varepsilon $ is
even faster for $b=20$ than for 
$b=4$. Nevertheless, the relative error stays below 7\% even for large precompressions,
so our approximation is still useful.

Finally, we point out that the values we have chosen for the precompression (up to 0.9)
correspond at most ($b = 20$) to $0.9 mm$ for the case studied in~\cite{herbold_solitary_2007}
(about 13\% of the diameter of the o-rings). Therefore, Fig.~\ref{fig:constpre} probably
encompasses most of the experimentally feasible cases. Above this value, the elastic
limit of the granular interaction would not hold.

\subsection{Gravitational precompression} 
\label{sub:gravpre}
For a vertical chain of granules, gravity causes an ever increasing
downward compression of the chain. That is, for the pair of granules $k$ and $k+1$, the
scaled force is
$f =  k {\cal G}$, where the scaled gravitational constant $\cal{G}$ was defined in
Eq.~(\ref{gravityscaling}).
We again rename the initial and maximum compression as
$z_{0,k}$ and $z_{m,k}$ to indicate explicitly that these quantities now vary along the chain and
hence depend on grain number $k$.  These compressions are now respectively the roots of the equations 
\begin{eqnarray}
    k {\cal G} &=& z_{0,k}^{3/2} + b z_{0,k}^{6}, \label{eq:z0-grav}\\
   4 k {\cal G} \left (  z_{m,k} - z_{0,k} \right ) &=& \frac{8}{5} \left (z_{m,k}^{5/2}- z_{0,k}^{5/2}\right ) \nonumber \\
  &&+ \frac{4}{7} b \left ( z_{0,k}^{7} - z_{m,k}^{7}\right ) - 1.\label{eq:zm-grav}
\end{eqnarray}
\begin{figure}
 \includegraphics[width=8cm]{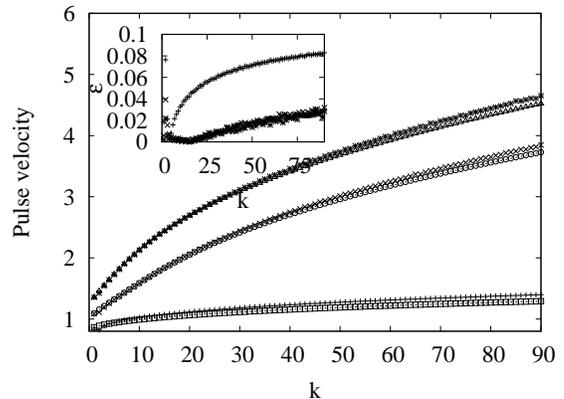}
 \caption{Pulse velocity as a function of granule number moving downward along the chain.
In the main figure, the plus signs (open squares), crosses (open circles) and stars (triangles)
refer to the binary collision approximation
for $b = 10^{-5}, \;4$ and $20$, respectively. The inset shows the relative error
$\varepsilon$ for the same values of $b$. \label{fig:gravpre}}
\end{figure}
Furthermore, the full potential also becomes $k$-dependent,
\begin{equation}
  V(k;z) = - 2 k {\cal G} z + \frac{4}{5} z^{5/2} + \frac{2}{7} b z^7.
\label{eq:potential-grav}
\end{equation}

Therefore, the residence time Eq.~(\ref{eq:residence}), and consequently the pulse
velocity, both become $k$-dependent as well. In Fig.~\ref{fig:gravpre} we plot the pulse
velocity as a function of the granule index. This dependence shows yet again
the good agreement between the binary collision approximation and the numerical
integration of the equations of motion. For this figure, we chose ${\cal G} = 0.024$.
This value is too large when compared with reasonable values for teflon o-rings. In fact,
for teflon o-rings the gravitational 
effects are negligible. However, it is interesting to understand the effects of gravity
since the theory is also applicable to other materials. The inset shows that
the error increases along the chain. This is not surprising, since the precompression
also increases along the chain. The velocity of granules along the chain increases
because of the increasing compression, and the binary collision approximation fails
to fully capture this increase.  Nevertheless, the effect of this failure is not very strong
for the experimental situations that have been tested, which tend to involve far fewer than the 90
granules that we have followed in our work.  Even for chains of up to 90 granules and even with
the exaggerated gravitational effects assumed here, the
relative error hardly exceeds 8\% in the worst case scenario (small $b$).

For single power-law materials, it has been shown that for large $k$ the
pulse velocity scales as $k^{[1-1/(n-1)]/2}$~\cite{sinkovits_nonlinear_1995,hong_power_1999}.
We have verified that even for our relatively short chains (short for the expectation of
scaling behavior), the pulse velocity is indeed a power law of $k$. For
$b=10^{-5}, \; c \sim k^{0.13},$ while for $b = 20, \; c \sim k^{0.36}$. These values
are not far from the values 0.17 and 0.42, respectively, which correspond to the limits of
a Hertzian or a power 7 potential. Hence, we can see that the binary
collision approximation already approaches the asymptotic behavior for chains of 90 granules. 

\section{Conclusion} 
\label{sec:conclusion}

The binary collision approximation has been very successful in
predicting pulse propagation behavior in granular chain, providing analytic results where only
numerical ones were previously available.
In this paper we have accomplished two goals in the further application of the binary collision
approximation to granular chains. One goal has been the extension of the method to chain in which
toroidal rings (o-rings) are placed between rigid cylinders that act as nonlinear springs, resulting
in potentials of interaction that contain two contributions rather than a single Hertzian one.  This
system is inspired by the availability of experimental results with which we can compare our
analytic outcomes.  The second, perhaps more important, goals is the extension of the binary
collision approximation methodology to chains with precompression, that is, beyond the sonic vacuum
cases considered in our earlier work.  This extension is a challenging test for the binary collision
approximation because precompression necessarily leads to situation in which more than two granules
participate substantially in each collision event.  When precompression is constant, the point of
eventual failure of the binary collision approximation must occur when the precompression force is
sufficiently strong.  In the case of gravitational precompression, failure must occur when the chain
is sufficiently long.  However, we find that for parameters that expansively cover experimental
regimes the binary collision approximation errs by relatively little.  The errors are of at most a
few percent when resulting pulse velocities are compared with those obtained by numerical
integration of the equations of motion of the full granular chain.  Given the differences in
the experimental and theoretical initial setups, the velocities predicted by the
binary collision approximation are gratifyingly close to the experimental values where the latter
are available.  We thus conclude that the binary collision approximation provides a powerful
analytic method for the study of pulse propagation in granular chain even in the presence of
precompression. We continue to examine the limits of applicability of this powerful methodology.

\section*{Acknowledgments}
Acknowledgment is made to the Donors of the American Chemical Society Petroleum Research Fund for
partial support of this research (KL).  AR acknowledges support from Bionanotec-CAPES
and CNPq.  AHR acknowledges support by CONACyT Mexico under Projects J-59853-F and J-83247-F.

\end{document}